\documentclass[aps,pra,groupedaddress,floatfix,twocolumn,showpacs,showkeys]{revtex4}

\usepackage{graphicx}
\usepackage{hyperref}
\usepackage{subfigure}

\usepackage{amsmath}
\usepackage{amsthm}
\usepackage{amssymb}

\newcommand{\eps}{\varepsilon}

\newcommand{\ket}[1]{\big|#1\big>}
\newcommand{\Ket}[1]{\left|#1\right>}

\newcommand{\set}[1]{\{#1\}}
\newcommand{\Set}[1]{\left\{#1\right\}}

\newcommand{\Norm}[1]{\left\|#1\right\|}

\newcommand{\abs}[1]{|#1|}
\newcommand{\Abs}[1]{\left|#1\right|}

\newcommand{\intervalcc}[2]{[#1,\, #2]}
\newcommand{\intervalco}[2]{[#1,\, #2)}

\DeclareMathOperator{\spanSpace}{span}

\newcommand{\ud}{\mathrm{d}}

\newtheorem{defi}{Definition}
\newtheorem{theo}[defi]{Theorem}

\newif\ifabsonly

\begin{document}

\title{Distributions of continuous-time quantum walks}

\date{\today}

\author{Arvid J. Bessen}
\email[]{bessen@cs.columbia.edu}
\affiliation{Columbia University\\
  Department of Computer Science%
  \footnote{Part of this work was done while the author was visiting the Institute of Quantum Computing and the University of Waterloo}}

\begin{abstract}
  We study the distributions of the continuous-time quantum walk on a one-dimensional lattice.
  In particular we will consider walks on unbounded lattices, walks with one and two boundaries and Dirichlet boundary conditions, and walks with periodic boundary conditions.
  We will prove that all continuous-time quantum walks can be written as a series of Bessel functions of the first kind and show how to approximate these series.
\end{abstract}

\pacs{02.50.-r, 03.67.Lx, 05.40.-a}

\maketitle

\section{Quantum Walks}

Quantum walks have been introduced as an algorithmic idea for quantum algorithms in an attempt to speed up classical random walk algorithms, see e.g. \cite{kem-03, amb-04}.
Two different proposals were made: the discrete-time quantum walk \cite{mey-96,aha-amb-kem-vaz-00} and the continuous-time quantum walk \cite{far-gut-98, chi-far-gut-02}.
While the behavior of the discrete-time quantum walk is well understood by now, the continuous-time quantum walk has not been studied as extensively, but see \cite{ger-wat-03, kon-04, got-04, str-05}.

In this paper we consider the one dimensional con\-tinuous-time quantum random walk on a line.
We write the state of the random walk system after time $t$ as
\begin{equation*}
  \Ket{\psi(t)} = \sum_{x = L}^{R} \psi(x,t) \Ket{x},
\end{equation*}
where $L, R \in \mathbb{Z} \cup \Set{\pm \infty}$, $L < R$, are the left and right boundaries of our walk.
Our goal is to determine $\psi(x,t)$ in an analytic form.
A continuous-time quantum random walk is now defined by the difference/differential equation
\begin{equation}\label{eqn:dif-dif-eqn}
  i \frac{\partial}{\partial t} \psi (x,t)
  =
  - \psi(x-1,t) + q \psi(x,t) - \psi(x+1,t) 
\end{equation}
and boundary conditions on $\psi(L,t), \psi(R,t)$.

In this paper we study four kind of boundary conditions:
the unbounded walk ($L= - \infty$ and $R = + \infty$),
walks with one boundary ($L$ finite, $R = + \infty$, and $\psi(L,t) = 0$),
walks with two boundaries ($L$, $R$ finite, $\psi(L,t) = \psi(R,t) = 0$),
and periodic boundary conditions ($L$, $R$ finite, $\psi(L,t) = \psi(R,t)$).

If both $L$ and $R$ are finite we define $N := R-L$ as the ``length'' of the interval in which the walk takes place.
In this case the continuous-time quantum walk from (\ref{eqn:dif-dif-eqn}) can also be written as
\begin{equation}\label{eqn:dif-dif-eqn-vec}
  i \frac{\partial}{\partial t} \Ket{\psi (t)}
  =
  - H_q \Ket{\psi(t)}.
\end{equation}
Here $H_q$ is a $(N-2) \times (N-2)$ matrix, when we use the $\psi(L,t) = \psi(R,t) = 0$ boundary conditions, or a $(N-1) \times (N-1)$ matrix for periodic boundary conditions, which is defined as
\begin{equation}\label{eqn:Hq}
  H_q =
  \begin{bmatrix}
    -q	& 1& & & \\
    1 & -q & 1 & & \\
    & \ddots & \ddots & \ddots & \\
    & & 1 & -q & 1 \\
    & & & 1 & -q \\
  \end{bmatrix}
\end{equation}
on the vector space $\spanSpace \Set{\Ket{L+1}, \Ket{L+2}, \ldots, \Ket{R-1}}$ for $\psi(L,t) = \psi(R,t) = 0$,
and
\begin{equation}\label{eqn:Hq-periodic}
  H_{q,p} =
  \begin{bmatrix}
    -q	& 1& & & 1 \\
    1 & -q & 1 & & \\
    & \ddots & \ddots & \ddots & \\
    & & 1 & -q & 1 \\
    1 & & & 1 & -q \\
  \end{bmatrix}
\end{equation}
on the vector space $\spanSpace \Set{\Ket{L}, \Ket{L+1}, \ldots, \Ket{R-1}}$ for periodic boundary conditions.

The solution $\ket{\psi(t)}$ of (\ref{eqn:dif-dif-eqn-vec}) for a given starting state $\ket{\psi(0)}$ is given by
\begin{equation}\label{eqn:evolution-eqn}
  \ket{\psi(t)} = e^{i H_{q} t} \ket{\psi(0)},
\end{equation}
which gives yet another definition of the continuous-time quantum random walk.

In this paper we will prove the following theorem:
\begin{theo}\label{theo:walk-functions}
  Let $J_n(t)$ denote the $n$-th Bessel function of the first kind, see e.g. \cite{abr-ste-72}. Define $\widetilde{J}_n(z) = i^n J_n (z)$.
  The continuous-time quantum random walk defined by equation (\ref{eqn:dif-dif-eqn}) with starting distribution $\Ket{\psi(0)} = \Ket{x_0}$ has the following coefficients $\psi(x,t)$ at time $t$:
  \begin{enumerate}
  \item For an unbounded walk, i.e. $L = - \infty$, $R = + \infty$,
    \begin{equation}\label{eqn:ctqrw-noboundary}
      \psi(x,t)
      =
      e^{- i t q} \widetilde{J}_{\abs{x-x_0}} (2t).
    \end{equation}
  \item For a left boundary $L \in \mathbb{Z}$ and $R = + \infty$ and boundary conditions $\psi(L,t) = 0$,
    \begin{equation}\label{eqn:ctqrw-oneboundary}
      \psi(x,t)
      =
      e^{- i t q}
      \left[
        \widetilde{J}_{\abs{x-x_0}} (2t)
        -
        \widetilde{J}_{\abs{x+x_0-2L}} (2t)
      \right]
      .
    \end{equation}
  \item
    For two boundaries $L, R \in \mathbb{Z}$, $L < R$ and boundary conditions $\psi(L,t) = \psi(R,t) = 0$,
    \begin{equation}\label{eqn:ctqrw-twoboundaries}
      \psi(x,t)
      =
      e^{- i t q}
      \sum_{n=-\infty}^{\infty}
      (-1)^n \widetilde{J}_{\abs{x-x_n}} (2t),
    \end{equation}
    with
    \begin{equation}\label{eqn:mirror-points}
      x_n :=
      \begin{cases}
        L-(x_{-n-1}-L) & n < 0 \\
        R+(R-x_{-n+1}) & n > 0
      \end{cases}
    \end{equation}
  \item
    For two boundaries $L, R \in \mathbb{Z}$, $L < R$ with periodic boundary conditions $\psi(L,t) = \psi(R,t)$,
    \begin{equation}\label{eqn:ctqrw-periodic}
      \psi(x,t)
      =
      e^{- i t q}
      \sum_{n=-\infty}^{\infty}
      \widetilde{J}_{\abs{x-y_n}} (2t),
    \end{equation}
    with
    \begin{equation}\label{eqn:periodic-points}
      y_n := n N + x_0 .
    \end{equation}
  \end{enumerate}
\end{theo}
The unbounded case (\ref{eqn:ctqrw-noboundary}) was known before for $q=0,2$, see \cite{far-gut-98, kon-04}.
For $q=0$ equations (\ref{eqn:ctqrw-oneboundary}) and (\ref{eqn:ctqrw-twoboundaries}) were derived by Apolloni and de Falco in their treatment of the clock of a quantum computer
\cite{apo-def-02}.
The walk with periodic boundary conditions (\ref{eqn:ctqrw-periodic}) was studied approximatively in \cite{gra-98,mul-blu-06}.
We will provide a unified proof of all these results in this paper.

The infinite series (\ref{eqn:ctqrw-twoboundaries}), (\ref{eqn:ctqrw-periodic}) are of course unwieldy for practical applications.
We can truncate (\ref{eqn:ctqrw-twoboundaries}) and (\ref{eqn:ctqrw-periodic}) after a certain number of terms and still get a very good approximation.
\begin{theo}\label{theo:approx}
  For two-sided and periodic boundary conditions the series (\ref{eqn:ctqrw-twoboundaries}) and (\ref{eqn:ctqrw-periodic}), truncated after
  \begin{equation*}
    k
    =
    \frac{t+N}{N}
    \frac{
      2 \ln t  + \frac{ \ln ( c \eps^{-1} t^{-1/2}) }{N}
    }{
      \ln (2 \ln t  + \frac{ \ln ( c \eps^{-1} t^{-1/2}) }{N})
    }
  \end{equation*}
  terms, with constant $c := \frac{22}{e \sqrt{2 \pi}}$, approximates $\psi(x,t)$ up to $\eps$, i.e.
  \begin{equation*}
    \Abs{
      \psi(x,t)
      -
      e^{- i t q}
      \sum_{n=-k}^{k}
      (-1)^n \widetilde{J}_{\abs{x-x_n}} (2t)
    }
    \leq
    \eps
  \end{equation*}
  for $0 < \eps < 1$ and $t > \max \set{1, \left( e^e (\eps/c)^{\frac{1}{N}}\right)^{\frac{2 N}{4 N - 1}} }$.
\end{theo}

\begin{figure*}[!htb]
\begin{center}
\subfigure[No boundaries]{
  \includegraphics[width=\columnwidth]{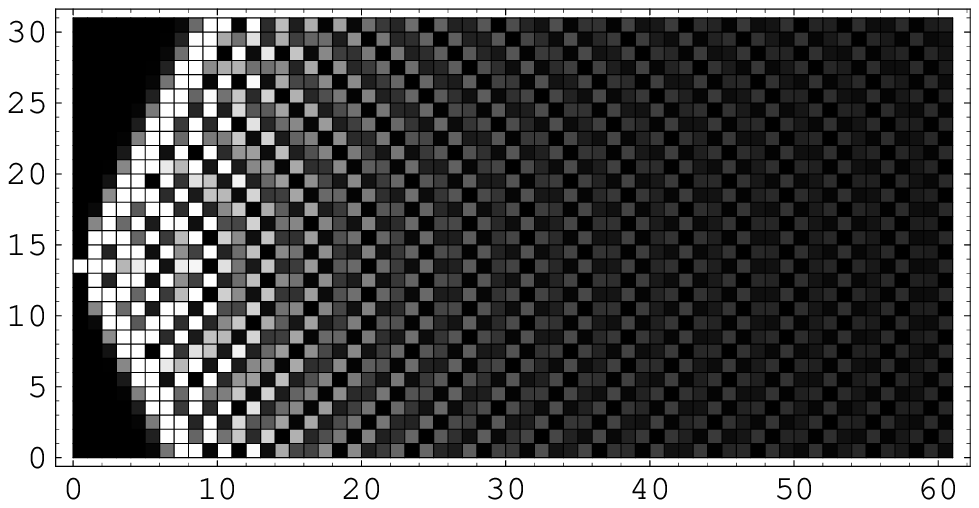}
}
\hfill
\subfigure[One boundary]{
  \includegraphics[width=\columnwidth]{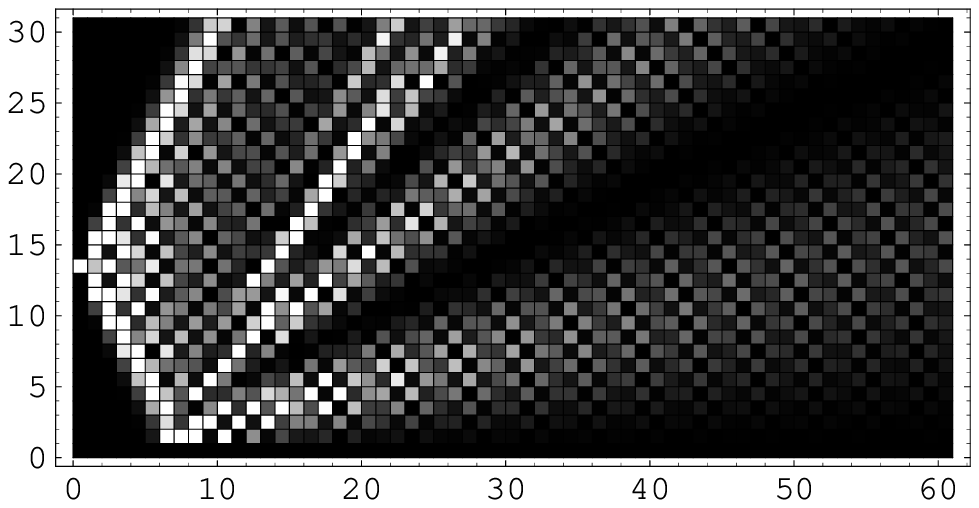}
}\\
\subfigure[Two boundaries]{
  \includegraphics[width=\columnwidth]{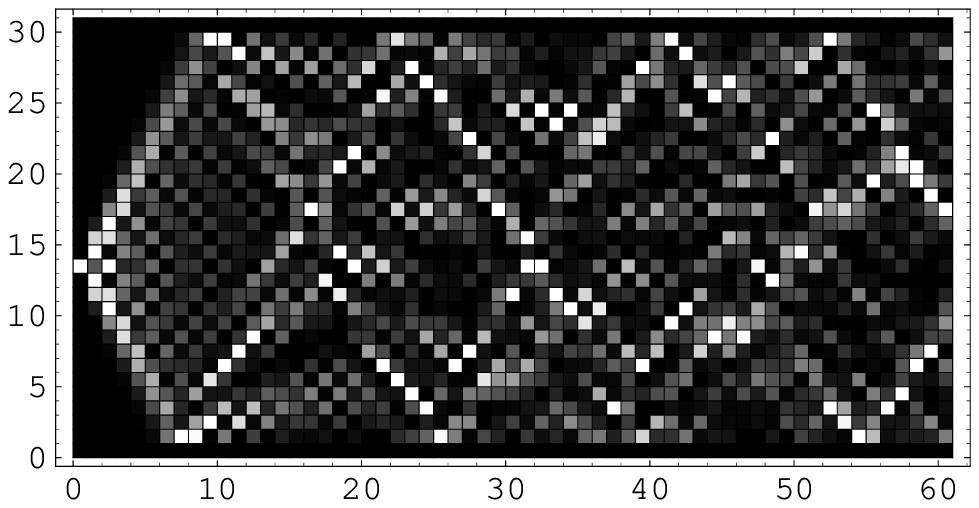}
}
\hfill
\subfigure[Periodic boundary conditions]{
  \includegraphics[width=\columnwidth]{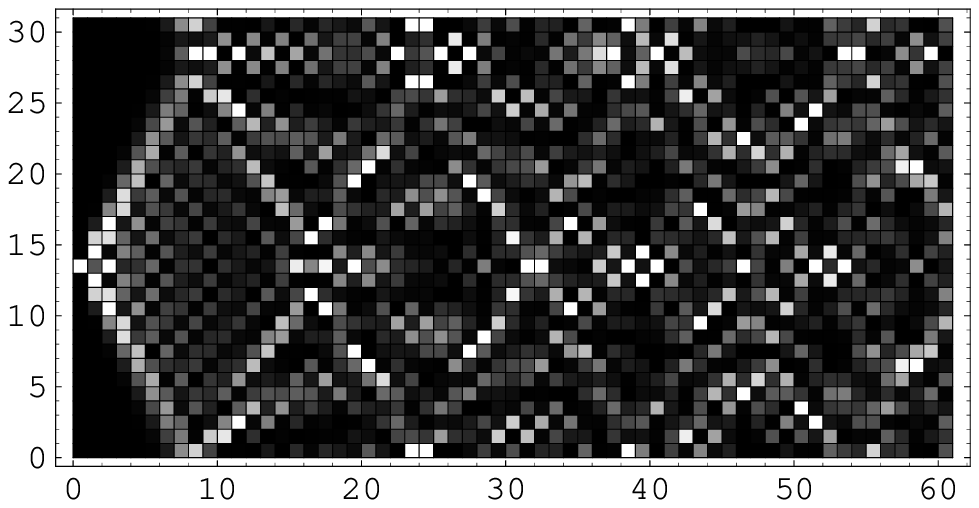}
}
\caption{\label{fig:dens-plots}Probabilities of a continuous-time quantum walk with $x_0 = 13$, $L=0$, $R=30$.
For two or periodic boundary conditions, we chose a precision of $10^{-5}$.
Therefore we can truncate after $k=12$ terms.}
\end{center}
\end{figure*}

We use Theorems \ref{theo:walk-functions}, \ref{theo:approx} to create a plot of the probability densities for a quantum walk.
In Figure \ref{fig:dens-plots} you see probability density plots that were made by truncating at the appropriate $k$ given by Theorem \ref{theo:approx}.

\ifabsonly
\else
\section{Analysis of walks with no boundaries and one boundary}

We can easily check that for $k=1, \ldots, N-1$
\begin{equation*}
  \Ket{\psi_k} = \sqrt{\frac{2}{N}} \sum_{x=L+1}^{R-1} \sin \left( \tfrac{k \pi (x-L)}{N} \right) \Ket{x}
\end{equation*}
are normalized and orthogonal eigenvectors of the matrix $H_q$ defined in (\ref{eqn:Hq}) with eigenvalues
$2 \cos \big( \tfrac{k \pi}{N} \big) - q$.
Therefore we can write $\Ket{\psi(t)} = e^{i t H_q} \Ket{\psi(0)}$ as
\begin{multline*}
  \psi(x,t)
  =
  \frac{2}{N}
  \sum_{k=1}^{N-1}
  \sin \big( \tfrac{\pi k (x-L)}{N} \big)
  e^{i t 2 \cos \big( \tfrac{k \pi}{N} \big) - i t q}
  \\
  \times
  \sum_{y=L+1}^{R-1}
  \sin \big( \tfrac{\pi k (y-L)}{N} \big)
  \psi(y,0)
  .
\end{multline*}
Suppose we start the random walk at $x_0$, i.e. our starting state $\Ket{\psi(0)} = \Ket{x_0}$ or equivalently $\psi(x,0) = \delta_{x,x_0}$.
Then
\begin{equation}\label{eqn:psixt}
  \begin{split}
    \psi(x,t)
    = &
    \frac{2 e^{- i t q}}{N}
    \sum_{k=1}^{N-1}
    \sin \big( \tfrac{\pi k (x-L)}{N} \big)
    \sin \big( \tfrac{\pi k (x_0-L)}{N} \big)
    \\
    & \ \hspace{40pt} \times
    e^{i t 2 \cos \big( \tfrac{k \pi}{N} \big)} \\
    = &
    \frac{e^{- i t q}}{N}
    \sum_{k=1}^{N-1}
    \cos \big( \tfrac{\pi k \abs{x-x_0}}{N} \big)
    e^{i t 2 \cos \big( \tfrac{k \pi}{N} \big) }
    \\
    &
    -
    \frac{e^{- i t q}}{N}
    \sum_{k=1}^{N-1}
    \cos \big( \tfrac{\pi k \abs{x+x_0-2L}}{N} \big)
    e^{i t 2 \cos \big( \tfrac{k \pi}{N} \big) } \\
  \end{split}
\end{equation}
We replace the sums with integrals.
The error for these approximations are (see e.g. the errors for Newton-Cotes quadrature rules in \cite{isa-kel-66}; note that the missing endpoints $k=0, N$ can be added without penalties, since they cancel out in (\ref{eqn:psixt})):
\begin{multline*}
  \Abs{
    \frac{1}{N}
    \left[
      \tfrac{1}{2} f(0)
      +
      \sum_{k=1}^{N-1}
      f(\tfrac{k}{N})
      +
      \tfrac{1}{2} f(1)
    \right]
    -
    \int_0^1 f(x) \ud x
  }
  \\
  \leq
  \mathcal{O} \left( \frac{1}{N^2} \Norm{ f''(x) }_{\infty} \right) ,
\end{multline*}
where the norm indicates the supremum of the second derivative of $f$ with respect to $x$ taken in the interval $\intervalcc{0}{1}$.
Since for $\alpha, \beta \geq 0$
\begin{equation*}
  \Norm{ \frac{\partial^2}{\partial w^2} \cos \big( \alpha w \big) e^{ i \beta \cos w} }_{\infty}
  \leq 
  (\alpha + \beta)^2 + \beta
\end{equation*}
we can rewrite equation (\ref{eqn:psixt}) as
\begin{equation*}
  \begin{split}
    \psi(x,t)
    = & \  
    e^{- i t q}
    \int_0^1 \cos ( \pi \abs{x-x_0} w )
    e^{ i t 2 \cos ( \pi w ) } \ud w \\
    & -
    e^{- i t q}
    \int_0^1 \cos ( \pi \abs{x+x_0-2L} w )
    e^{i t 2 \cos ( \pi w ) } \ud w \\
    & \pm
    e(x,x_0,t,N)
  \end{split}
\end{equation*}
with an error term of order
\begin{equation}\label{eqn:error}
  e(x,x_0,t,N)
  =
  \mathcal{O}
  \left(
    \tfrac{(\abs{x-x_0}+2t)^2 + (\abs{x+x_0-2L}+2t)^2}{N^2}
  \right)
  .
\end{equation}
This representation allows us to express $\psi(x,t)$ as a sum of Bessel functions.
According to \cite{abr-ste-72}, equation 9.1.21, for $n \in \mathbb{N}^{+}$ the Bessel function $J_n(z)$ is given by
\begin{equation*}
  J_n(z) = \frac{i^{-n}}{\pi} \int_{0}^{\pi} e^{i z \cos w} \cos (n w) \ud w
\end{equation*}
and therefore
$
\widetilde{J}_n(z) = \frac{1}{\pi} \int_{0}^{\pi} e^{i z \cos w} \cos (n w) \ud w
$,
and we can write
\begin{equation}\label{eqn:psixt-bessel}
  \begin{split}
    \psi(x,t)
    = & \ 
    e^{- i t q}
    \left[
      \widetilde{J}_{\abs{x-x_0}} (2t)
      - 
      \widetilde{J}_{\abs{x+x_0-2L}}(2t)
    \right]
    \\
    & \
    \pm e(x,x_0,t,N) .
  \end{split}
\end{equation}
For $R \rightarrow \infty$, which implies $N = R-L \rightarrow \infty$, the error term $e(x,x_0,t,N)$ defined in (\ref{eqn:error}) converges to $0$ and we get
\begin{equation*}
  \psi(x,t)
  =
  e^{- i t q}
  \left[
    \widetilde{J}_{\abs{x-x_0}} (2t)
    - 
    \widetilde{J}_{\abs{x+x_0-2L}}(2t)
  \right]
  , \\
\end{equation*}
which proves (\ref{eqn:ctqrw-oneboundary}).
This is the limiting distribution for a continuous-time quantum random walk with one boundary.
For no boundaries, i.e. if we let $L \rightarrow \infty$, too, (\ref{eqn:ctqrw-oneboundary}) simplifies to
\begin{equation*}
  \psi(x,t)
  =
  e^{- i t q} \widetilde{J}_{\abs{x-x_0}} (2t) ,
\end{equation*}
since $\lim_{n \rightarrow \infty} J_n(x) = 0$, which proves (\ref{eqn:ctqrw-oneboundary}) and can be compared to the results from \cite{far-gut-98,kon-04}.

\section{Walks with two boundaries}

To derive the correct form of the quantum walk distribution for walks with two boundaries we check the correctness of Equation (\ref{eqn:ctqrw-noboundary}) for the unbounded case by plugging it into (\ref{eqn:dif-dif-eqn}):
\begin{multline*}
  i \frac{\partial}{\partial t} \psi (x,t) 
  =
  i \frac{\partial}{\partial t} \left[ i^{\Abs{x-x_0}} e^{- i t q} J_{\Abs{x-x_0}} (2t) \right] \\
  = 
  i i^{\abs{x-x_0}}
  \Big[
  - i q
  e^{- i t q}
  J_{\abs{x-x_0}} (2t)
  +
  e^{- i t q}
  J'_{\abs{x-x_0}} (2t)
  \Big]
  ,
\end{multline*}
where we use that the derivative of the Bessel function satisfies $J'_{\nu} (z) = \frac{J_{\nu-1}(z) - J_{\nu+1}(z)}{2}$, see e.g. \cite{abr-ste-72}, to verify that $\psi $ indeed fulfills (\ref{eqn:dif-dif-eqn}), at least for $\abs{x-x_0} \geq 1$.
In the special case $\abs{x-x_0} = 0$ we use that $J'_0(z) = - J_1(z)$.

To satisfy the boundary conditions $\psi(L,t) = \psi(R,t) = 0$ we use the method of images, see \cite{mor-fes-53}, and introduce mirror images of the walk starting at $x_0$.
Let $\ket{\psi_{x_0} (t)}$ be the solution for an unbounded walk starting at $x_0$.
Then $\ket{\psi^{(0)}_{x_0} (t)} = \ket{\psi_{x_0} (t)}$ is a good approximation to the solution, but we have to correct the error on the boundary by introducing walks starting at $x_{-1} = L-(x_0-L)$ and $x_1 = R+(R-x_0)$:
\begin{equation*}
  \ket{\psi^{(1)}_{x_0}(t)}
  =
  \ket{\psi^{(0)}_{x_0}(t)}
  - \ket{\psi_{x_{-1}}(t)}
  - \ket{\psi_{x_1}(t)} .
\end{equation*}
To correct the error introduced by the corrections we continue to add mirror images at $x_{-2} =  L-(x_1-L)$ and $x_2 = R+(R-x_{-1})$, ..., until we arrive at (\ref{eqn:ctqrw-twoboundaries}).
Let us check that
\begin{equation*}
  \begin{split}
    \frac{\psi(L,t)}{e^{- i t q}}
    & = 
    \sum_{n=-\infty}^{\infty}
    (-1)^n \widetilde{J}_{\abs{L-x_n}} (2t)
    \\
    &
    = 
    \sum_{n=0}^{\infty}
    (-1)^n \left[
      \widetilde{J}_{\abs{L-x_n}} (2t)
      - \widetilde{J}_{\abs{L-x_{-(n+1)}}} (2t)
    \right] \\
    & = 
    \sum_{n=0}^{\infty}
    (-1)^n \left[
      \widetilde{J}_{\abs{L-x_n}} (2t)
      - \widetilde{J}_{\abs{x_n-L}} (2t)
    \right]
    = 0,
  \end{split}
\end{equation*}
and similarly $\psi(R,t) = 0$.
This proves (\ref{eqn:ctqrw-twoboundaries}).

For periodic boundary conditions $\psi(L,t) = \psi(R,t)$ we also build a solution starting from $\Ket{\psi_{x_0}(t)}$.
In order to enforce periodicity, we add two walks at $y_{-1} = x_0 - N$ and $y_1 = x_0 + N$, which is the correction term for the ``first round''.
If we define
$$
y_n = 
x_0 + n N
$$
we arrive at (\ref{eqn:ctqrw-periodic}).
Then
\begin{equation*}
  \begin{split}
    \frac{\psi(L,t)}{e^{- i t q}}
    &
    = 
    \sum_{n=-\infty}^{\infty}
    \widetilde{J}_{\abs{L-y_n}} (2t)
    = 
    \sum_{n=-\infty}^{\infty}
    \widetilde{J}_{\abs{L - n N - x_0}} (2t)
    \\
    &
    = 
    \sum_{n=-\infty}^{\infty}
    \widetilde{J}_{\abs{R - (n+1) N - x_0}} (2t)
    =
    \frac{\psi(R,t)}{e^{- i t q}} ,
  \end{split}
\end{equation*}
which completes the proof of Theorem \ref{theo:walk-functions}.

\section{Approximations}

For practical purposes the infinite series in Theorem \ref{theo:walk-functions} are unsatisfactory.
Theorem \ref{theo:approx}, which we will prove in this section, fortunately tells us that we can truncate those series and still get an excellent approximation to the correct distribution.

Consider the distribution $\psi(x,t)$ in (\ref{eqn:ctqrw-twoboundaries}) for zero boundary conditions
    \begin{equation*}
      \psi(x,t)
      =
      e^{- i t q}
      \sum_{n=-\infty}^{\infty}
      (-1)^n \widetilde{J}_{\abs{x-x_n}} (2t),
    \end{equation*}
and truncate it after the first $k$ terms.
We would like to determine the error of this approximation, i.e.
    \begin{equation*}
      e_{k,0}
      =
      \Abs{
      \psi(x,t)
      -
      e^{- i t q}
      \sum_{n=-k}^{k}
      (-1)^n \widetilde{J}_{\abs{x-x_n}} (2t)
      } ,
    \end{equation*}
for $t \geq 0$.
We can bound the Bessel functions $J_{\nu}(2t)$ by
\begin{equation}\label{eqn:bessel-upper}
  \Abs{J_{\nu}(2t)}
  \leq
  \frac{\Abs{t}^{\nu}}{\Gamma(\nu+1)} ,
\end{equation}
see \cite{abr-ste-72}.
Then the error $e_{k,0}$ is bounded by
\begin{equation*}
\begin{split}
  e_{k,0} \leq &
  \sum_{n=k+1}^{\infty}
  \Abs{\widetilde{J}_{\abs{x-x_n}} (2t)}
  +
  \sum_{n=-k-1}^{-\infty}
  \Abs{\widetilde{J}_{\abs{x-x_n}} (2t)} \\
  \leq &
  \sum_{n=k+1}^{\infty}
  \left[
  \frac{t^{\abs{x-x_n}}}{\Gamma(\abs{x-x_n}+1)}
  +
  \frac{t^{\abs{x-x_{-n}}}}{\Gamma(\abs{x-x_{-n}}+1)}
  \right] .
\end{split}
\end{equation*}
We similarly define the $k$ term approximation error $e_{k,p}$ for periodic boundary conditions.

From the definition (\ref{eqn:mirror-points}) of the reflection points $x_n$
\begin{equation*}
  x_n =
  \begin{cases}
    L-(x_{-n-1}-L) & n < 0 \\
    R+(R-x_{-n+1}) & n > 0
  \end{cases}
\end{equation*}
we see that $L + n N \leq x_n \leq R + n N$.
Therefore, and since $L \leq x \leq R$,
\begin{equation}\label{eqn:ek0}
\begin{split}
  e_{k,0} \leq &
  \sum_{n=k+1}^{\infty}
  \left[
  \frac{t^{\abs{x-x_n}}}{\Gamma(\abs{x-x_n}+1)}
  +
  \frac{t^{\abs{x-x_{-n}}}}{\Gamma(\abs{x-x_{-n}}+1)}
  \right] \\
  \leq &
  \sum_{n=k+1}^{\infty}
  \left[
  \frac{t^{(n+1) N}}{\Gamma((n-1)N+1)}
  +
  \frac{t^{(n+1) N}}{\Gamma((n-1)N+1)}
  \right] \\
  \leq &
  2 t^{N}
  \sum_{n=k+1}^{\infty}
  \frac{t^{n N}}{((n-1) N)!}
  =
  2 t^{2N}
  \sum_{n=k}^{\infty}
  \frac{t^{n N}}{(n N)!} .
\end{split}
\end{equation}
The same error estimate also holds for the error $e_{k,p}$ of the $k$-term approximation to the series (\ref{eqn:ctqrw-periodic}) for periodic boundary conditions.

To estimate $e_{k,0}$ and $e_{k,p}$ in (\ref{eqn:ek0}) we can use Stirling's formula
\begin{equation*}
  x!
  =
  \sqrt{2 \pi} x^{x+\frac{1}{2}} e^{-x+\frac{\theta}{12 x}}
  \geq
  \sqrt{2 \pi} x^{x+\frac{1}{2}} e^{-x},
\end{equation*}
see \cite{abr-ste-72}, where $x>0$ and $0 < \theta < 1$.
Then
\begin{equation*}
  \begin{split}
    ( n N)!
    \geq &
    \sqrt{2 \pi} (n N)^{( n N)+\frac{1}{2}} e^{-( n N)} \\
    \geq &
    \sqrt{2 \pi n N} (n N e^{-1})^{n N} .
  \end{split}
\end{equation*}
Therefore the error $e_{k,0}$ from (\ref{eqn:ek0}) is bounded by
\begin{equation*}
    e_{k,0}
    \leq
    \frac{2 t^{2N}}{\sqrt{2 \pi N}}
    \sum_{n=k}^{\infty}
    \frac{1}{\sqrt{n}}
    \left[ \frac{t e}{n N} \right]^{n N}
    \hspace{-3pt}
    \leq
    \frac{2 t^{2N}}{\sqrt{2 \pi N k}}
    \sum_{n=k}^{\infty}
    \left[ \frac{t e}{n N} \right]^{n N}
    \hspace{-3pt}
    .
\end{equation*}
This series converges if $k = k(t, \eps)$ is chosen such that 
$k > \frac{t e}{N}$.
In this case
\begin{equation}\label{eqn:tricky-ineq}
  \begin{split}
    \sum_{n=k}^{\infty}
    \left( \frac{t e}{n N} \right)^{N n}
    \leq &
    \left( \frac{t e}{k N} \right)^{N k}
    \sum_{n=k}^{\infty}
    \left( \frac{t e}{k N} \right)^{N (n-k)} \\
    = &
    \left( \frac{t e}{k N} \right)^{N k}
    \frac{1}{1 - \left( \frac{t e}{k N} \right)^{N}} \\
    = &
    \frac{e^{- N k \ln \frac{k N}{t e}}}{1-\left( \frac{t e}{k N} \right)^{N}}
    .
  \end{split}
\end{equation}
Therefore the error from truncating after $k$ terms is
\begin{equation}\label{eqn:ek-nosums}
  \begin{split}
    e_{k,0}
    \leq &
    \frac{2 t^{2N}}{\sqrt{2 \pi e N k}}
    \frac{e^{- N k \ln \frac{k N}{t e}}}{1 - \left( \frac{t e}{k N} \right)^{N}}
  \end{split}
\end{equation}

We want to use (\ref{eqn:ek-nosums}) to find $k$ such that $e_{k,0} = \mathcal{O}( \eps )$.
We make the following ansatz
\begin{align}\label{eqn:k-ansatz}
  k
  = &
  k(t,\eps)
  =
  \frac{t+N}{N}
  \frac{\zeta}{\ln \zeta},
  \text{ where}
  \\
  \label{eqn:zeta}
  \zeta
  = &
  \zeta(t,\eps) =
  2 \ln t  + \frac{\ln (c \, \eps^{-1} t^{-1/2})}{N}
\end{align}
with a constant $c > 0$ that we will determine.
We have to check
that the requirement $k > \frac{t e}{N}$ that guarantees convergence in (\ref{eqn:tricky-ineq}) is fulfilled.
If we restrict $t$ to
\begin{equation}\label{eqn:t-req}
  t \geq \left( e^e (\eps/c)^{\frac{1}{N}}\right)^{\frac{2 N}{4 N - 1}}
\end{equation}
then
\begin{equation}\label{eqn:zeta-lower}
  \zeta
  =
  2 \ln t  + \frac{\ln (c \, \eps^{-1} t^{-1/2})}{N}
    \geq
    e .
\end{equation}
Additionally we observe
that $\frac{z}{\ln z}$ is growing mo\-no\-to\-nous\-ly for $z > e$,
\begin{equation*}
  \frac{\ud}{\ud z} \frac{z}{\ln z} = \frac{\ln z - 1}{(\ln z)^2} > 0
  \text{ for } z > e,
\end{equation*}
and therefore for $t$ chosen as in (\ref{eqn:t-req})
\begin{equation}\label{eqn:k-est}
\begin{split}
  k
  = &
  \frac{t+N}{N}
  \frac{\zeta}{\ln \zeta}
  \geq
  \frac{t+N}{N}
  e
  >
  \frac{t e}{N} .
\end{split}
\end{equation}

We estimate the numerator and denominator of $e_{k,0}$ in (\ref{eqn:ek-nosums}) separately and write $e_{k,0} \leq \frac{M(k)}{D(k)}$.
If we insert (\ref{eqn:k-ansatz})
into the numerator of  (\ref{eqn:ek-nosums}), we see that
\begin{equation*}
  \begin{split}
    M(k)
    := &
    2 t^{2N}
    e^{- N k \ln \frac{kN}{t e} }
    =    
    2 t^{2N}
    e^{- N \frac{t+N}{N} \frac{\zeta}{\ln \zeta} \ln \left( \frac{t+N}{N} \frac{\zeta}{\ln \zeta} \frac{N}{t e} \right) } \\
    = &
    2 t^{2N}
    e^{
      - (t+N) \zeta
      \left( 1 + \frac{\ln \frac{t+N}{t e} - \ln \ln \zeta}{\ln \zeta} \right)
    } .
  \end{split}
\end{equation*}
Now we use that we can show that for $z > 1$
\begin{equation*}
  \frac{\ln \frac{t+N}{t e} - \ln \ln z}{\ln z}
  \geq
  \frac{\ln \frac{t+N}{t e} - \ln \ln e^{\frac{t+N}{t}}}{\ln e^{\frac{t+N}{t}}}
  =
  - \frac{t}{t+N} .
\end{equation*}
We can use this to estimate $M(k)$ since $\zeta \geq e > 1$, see (\ref{eqn:zeta-lower}), and get
\begin{equation*}
  M(k)
  \leq
  2 t^{2N}
  e^{- \zeta N}
  =
  2 t^{2N}
  e^{- 2 N \ln t}
  e^{\ln (c^{-1} \eps t^{1/2})}
  =
  \frac{2}{c} \eps \sqrt{t} .
\end{equation*}

To estimate the denominator of (\ref{eqn:ek-nosums})
\begin{equation*}
  D(k) :=
  \sqrt{2 \pi e N k}
  \left( 1 - \left( \frac{t e}{k N} \right)^{N} \right)
\end{equation*}
we insert (\ref{eqn:k-ansatz}) in
\begin{equation}\label{eqn:denom-base}
   \frac{t e}{k N}
   =
   \frac{t e}{\frac{t+N}{N} \frac{\zeta}{\ln \zeta} N}
   =
   \frac{t}{t+N}
   \frac{e \ln \zeta}{ \zeta }
   .
\end{equation}
We can see that both $\frac{t}{t+N}$ and $\frac{e \ln \zeta}{\zeta}$ are less than $1$ for $t$ as in (\ref{eqn:t-req}), when we have $\zeta \geq e$, see (\ref{eqn:zeta-lower}).
We analyze (\ref{eqn:denom-base}) by considering $t \in \intervalco{( e^e (\eps/c)^{\frac{1}{N}} )^{\frac{2 N}{4 N - 1}}}{\infty}$ over two intervals.
For $t \in \intervalco{( e^e (\eps/c)^{\frac{1}{N}} )^{\frac{2 N}{4 N - 1}}}{10}$
\begin{equation*}
   \frac{t}{t+N}
   \frac{e \ln \zeta}{ \zeta }
   \leq
   \frac{t}{t+N}
   \leq
   \frac{t}{t+1}
   \leq
   \frac{10}{11}
   .
\end{equation*}
In the interval $\intervalco{10}{\infty}$ we use that $\frac{e \ln \zeta}{\zeta}$ converges monotonously to $0$ for $\zeta = \zeta(t,\eps) \rightarrow \infty$, and that $\zeta \geq 2 \ln t$:
\begin{equation*}
   \frac{t}{t+N}
   \frac{e \ln \zeta}{ \zeta }
   \leq
   \frac{e \ln \zeta}{ \zeta }
   \leq
   \frac{e \ln (2 \ln t)}{ 2 \ln t }
   \leq
   0.902
   \leq
   \frac{10}{11}
   .
\end{equation*}
Therefore we can use this and (\ref{eqn:k-est}) for
\begin{equation*}
  D(k)
  \geq
  \sqrt{2 \pi e N k}
  \frac{1}{11}
  >
  \frac{e \sqrt{2 \pi t}}
  {11}
  .
\end{equation*}

If we combine the results for numerator and denominator, we can estimate
\begin{equation*}
  e_{k,0} \leq \frac{M(k)}{D(k)} \leq \frac{22}{c e \sqrt{2 \pi}} \eps,
\end{equation*}
and we see that we have to choose $c := \frac{22}{e \sqrt{2 \pi}}$.
This result holds for all $t$ chosen according to (\ref{eqn:t-req}).

If we follow the same steps we can also estimate the error $e_{k,p}$ of a $k$ term approximation for periodic boundary conditions.
This proves Theorem \ref{theo:approx}.

\fi 

\section{Acknowledgments}

The author would like to thank the members of the Institute for Quantum Computing and the University of Waterloo for their hospitality and many intriguing discussions.
Special thanks to Michele Mosca, Hilary Carteret and Bruce Richmond for helpful comments.
The author would also like to thank David DiVincenzo for pointing out the description of the method of images in \cite{mor-fes-53} to him and Henryk Wo{\'z}niakowkski for his advice regarding the convergence rate of the power series.

\bibliographystyle{abbrv}
\bibliography{../qc}

\end{document}